\documentclass[preprint,12pt]{elsarticle}
\usepackage{listings}
\usepackage[usenames,dvipsnames]{color}
\usepackage{multirow}
\usepackage{soul}

\usepackage{epic}
\usepackage{eepic}
\usepackage{makeidx}
\usepackage{array}
\usepackage{times}
\usepackage{amssymb}
\usepackage{amsmath}
\usepackage{amsxtra}
\usepackage{graphicx}
\usepackage{subfig}
\usepackage{boxedminipage}
\usepackage{alltt}
\usepackage{tikz}
\usepackage{booktabs}
\usepackage{xspace}
\usepackage[pdftex=true,hyperindex=true,colorlinks=true,breaklinks=true, urlcolor=red, linkcolor=blue, citecolor=black,bookmarksnumbered=true,pdfpagelabels=true,plainpages=false]{hyperref}
\usepackage{rotating}
\usepackage{float}
\usepackage{listings}
\usepackage{url}

\newcommand{\add}[1]{#1}

\def\midtilde{\raise-0.6ex\hbox{\textasciitilde}}

\journal{Computer Physics Communications}

\begin{document}

\definecolor{sand}{RGB}{193,154,107} 
\def\cols{20}                        
\def\rows{40}                        
\def\SquareUnit{.35}                 
\pgfmathsetmacro\RmaxParticle{.1}    
\def\BeforeLight{5}                  

\begin{frontmatter}

\title{Real-Time Computation of Parameter Fitting and Image Reconstruction Using Graphical Processing Units}

\author[label1,label3]{Uldis Locans}
\author[label1]{Andreas Adelmann\corref{mycorrespondingauthor}}
\cortext[mycorrespondingauthor]{Corresponding author}
\ead{andreas.adelmann@psi.ch}

\author[label1]{Andreas Suter}
\author[label2]{Jannis Fischer}
\author[label2]{Werner Lustermann}
\author[label2]{G\"unther Dissertori}
\author[label2,label4]{Qiulin Wang}

\address[label1]{Paul Scherrer Institut, Villigen, CH-5232, Switzerland}
\address[label2]{ETH Institut f\"ur Teilchenphysik, CH-8092, Switzerland}
\address[label3]{University of Latvia, 19 Raina Blvd., Riga, LV 1586, Latvia}
\address[label4]{Tsinghua University Institute of Medical Physics, Bejing, China}



\begin{abstract}



In recent years graphical processing units (GPUs) have become a powerful tool in scientific computing. Their potential to speed up highly parallel applications brings the power of high performance computing to a wider range of users. However, programming these devices and integrating their use in existing applications is still a challenging task.

In this paper we examined the potential of GPUs for two different applications. The first application, created at Paul Scherrer Institut (PSI), is used for parameter fitting during data analysis of $\mu$SR (muon spin rotation, relaxation and resonance) experiments. The second application, developed at ETH, is used for PET (Positron Emission Tomography) image reconstruction and analysis. Applications currently in use were examined to identify parts of the algorithms in need of optimization. Efficient GPU kernels were created in order to allow applications to use a GPU, to speed up the previously identified parts. Benchmarking tests were performed in order to measure the achieved speedup. 

During this work, we focused on single GPU systems to show that real time data analysis of these problems can be achieved without the need for large computing clusters. The results show that the currently used application for parameter fitting, which uses OpenMP to parallelize calculations over multiple CPU cores, can be accelerated around 40 times through the use of a GPU. The speedup may vary depending on the size and complexity of the problem. For PET image analysis, the obtained speedups of the GPU version was more than $\times$40 larger compared to a single core CPU implementation. The achieved results show that it is possible to improve the execution time by orders of magnitude.

\end{abstract}


\end{frontmatter}

\section{Introduction}

The use of graphical processing units (GPUs) in scientific applications has been increasing in recent years. With the emergence of development frameworks such as CUDA and OpenCL, the computational power of GPUs is becoming more available for use in general purpose and scientific applications. The capabilities of GPUs to increase the performance of highly parallel algorithms, and the relatively low price of these devices, allows the creation of cost effective real time simulation and data analysis systems.

This paper describes the efforts to use the power of GPUs to achieve near real time performance for data analysis of two separate problems. The first application that this paper describes focuses on the parameter fitting for data analysis of $\mu$SR experiments. The second application performs PET image reconstruction and analysis.

Parameter fitting using \textsc{Musrfit} is examined in this paper. This application is used for data analysis of $\mu$SR experiments. During the fitting, a set of parameters is determined by performing $\chi^2$ minimization. The most time consuming part of this parameter fitting is the calculation of the $\chi^2$ values. Using the current CPU implementation, this fitting can take hours for certain data sets. The calculation of $\chi^2$ is a good algorithm to parallelize using a GPU and the approach used to create a real time parameter fitting using a GPU is described in this paper.

The second part of the paper describes the efforts to speed up an iterative algorithm for tomographic image reconstruction. In addition to reconstruction, this application also performs an analysis of the reconstructed image which searches for signals above a background.

When integrating GPU usage in existing applications, it can add another level of complexity to the program code. For this reason the dynamic kernel scheduler (DKS) was developed and described in \cite{dks}. DKS allows the separation of all the device code details into an independent layer and provides a simple interface for the host application to invoke tasks on the GPU or other accelerator devices. DKS was also used in this work to keep the \textsc{Musrfit} and the image recognition GPU code separate from the original application. This improves the code maintenance and keeps the application more portable, as it is possible to disable the DKS provided layer if there is no GPU device available on the system.

For parameter fitting the state-of-the-art application for data analysis of $\mu$SR experiments was chosen as the baseline implementation. The application uses OpenMP to parallelize the calculation of $\chi^2$ on the CPU. For image reconstruction and analysis the application used at ETH for analysis of experimental data was chosen as the baseline. The algorithm implementation in this application is serial and uses just one core of the CPU. The main aim of this work was to show how both of these applications could benefit from the use of hardware accelerators such as GPUs. By the use of DKS, that provides a higher level of abstraction from the host application side, we demonstrate seamless and simple integration of the GPU code in the existing software. The higher level of abstraction will provide us with software investment protection because one can react prompt and effortless on hardware changes on the GPU side. The simplicity will help maintaining software quality
in our inhomogeneous (w.r.t. software engineering competences) open source development team.

The rest of the paper is organized as follows: Section \ref{sec:relatedwork} describes related work, section \ref{sec:dks} provides an overview of DKS and how it was used in regards to the work presented. Section \ref{sec:musrfit} provides a description of: the application that is being used for parameter fitting, the problems the application is applied to, and the efforts to improve the performance of this application with the use of a GPU. At the end of the Section \ref{sec:musrfit}, benchmark results of a specific parameter fitting example are presented and analyzed. Section \ref{sec:recon} provides: insight into the image reconstruction and analysis problem, the specifics of how GPU was used to speed up the algorithms, and the details on the achieved results with GPUs. In Section \ref{sec:conclusions} conclusions are presented.

\section{Related work}
\label{sec:relatedwork}
With the increasing popularity of hardware accelerators there have been many attempts to provide high level APIs that allow the creating of GPU code. Nvidia provides a set of GPU accelerated libraries with the CUDA toolkit \mbox{\cite{cudalibraries}} that can be easily incorporated in the host applications. Several parallel vector libraries, such as Thrust \mbox{\cite{thrust}}, ArrayFire \mbox{\cite{arrayfire}} and BoostCompute \mbox{\cite{boostcompute}}, are available that implement parallel versions of algorithms from C++ standard template libraries. There have also been attempts to create higher level APIs and abstractions to ease the creation of the GPU code \mbox{\cite{Bourgoin2014,Svensson2010,Vinas2015}}. These attempts focus on creating a more generic way of expressing the GPU code that is later translated to CUDA or OpenCL kernels.

The DKS API does not aim to replace or replicate these efforts but rather provide a confined layer where all of these approaches can be used, and together with hand tuned kernels provides the ability to create fast, optimized algorithms for hardware accelerators. DKS algorithms for parameter fitting use handwritten kernels complemented by cuBLAS library, while image analysis in addition to hand written kernels uses Thrust libraries sort functions.

\section{Dynamic Kernel Scheduler}
\label{sec:dks}
With DKS we aim to add GPUs to complex physics applications which previously only relied upon the power of the CPU for computation.
DKS is a software module that allows to move all of the device specific code in a separate layer and provides a simple interface that can be used in the existing host application to offload tasks to the GPU. This approach eases the integration of hardware accelerators in existing scientific applications while still allowing to create fast, optimized kernels to run on the GPU.

CUDA and OpenCL kernels used in DKS are hand tuned and optimized for optimal performance. In future versions of DKS an auto-tuning module is planned, that would allow to optimize kernel launch parameters, but it is not included in current work.

\begin{figure}[ht]
\centering
\includegraphics[height=3in, width=3in]{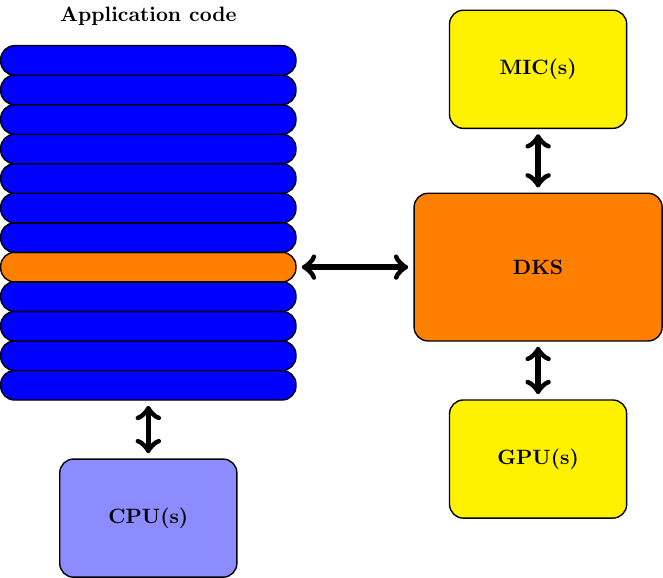}
\caption{Dynamic Kernel Scheduler concept.}
\label{fig:dks_concept}
\end{figure}

The concept of how DKS and host application work together is shown in Figure \ref{fig:dks_concept}.

\noindent The algorithms that are offloaded to the GPU are implemented in the DKS. The host application uses a simple interface to communicate the tasks that need to be executed on the device, and providing the necessary data. 

As an example, a Fast Fourier Transformation offload using DKS is shown in Code sample \ref{code:dksexample}. The simplicity and high level of abstraction is evidently shown, in this simple but representative template of a generic offloading task.

\begin{figure}[ht]
 \centering
\begin{lstlisting}[frame=single, caption={Example of DKS interface integrated in the host application.}, label={code:dksexample}]
  //setup and initialize the device
  DKSbase dks;
  dks.setAPI("CUDA");
  dks.setDevice("-gpu");
  dks.initDevice();

  //allocate memory on device and write data
  mem_ptr = dks.allocateMemory<Complex_t>(DATA_SIZE);
  dks.writeData<Complex_t>(mem_ptr, DATA_ARRAY, DATA_SIZE);

  //execute FFT or IFFT
  if (direction == 1)
    dks.callFFT(mem_ptr, DIMENSIONS, DIM_SIZE);
  else
    dks.callIFFT(mem_ptr, DIMENSIONS, DIM_SIZE);

  //read data and free memory
  dks.readData<Complex_t>(mem_ptr, DATA_ARRAY, DATA_SIZE);
  dks.freeMemory<Complex_t>(mem_ptr, DATA_SIZE);
}
\end{lstlisting}
\end{figure}

The separation of device code in a different module makes both the host application and GPU code a lot more manageable and maintainable. \add{This adds some software investment protection since the algorithms in DKS can be improved and updated using the best available tools and libraries to ensure the best possible performance with no changes to the host application.} It also opens the possibility to add different devices in the future with no or very few modifications to the host application, since with DKS it is possible to create the device code using multiple frameworks such as CUDA, OpenCL, or OpenMP. During this work we focused on the use of CUDA to take advantage of computational power of NVIDIA GPUs. For parameter fitting, OpenCL was used as well to make the application more flexible and capable of running on other devices.

\section{Parameter fitting with Musrfit}
\label{sec:musrfit}

\textsc{Musrfit} is a software tool for analyzing time-differential $\mu$SR data \cite{suter2012musrfit}. \textsc{Musrfit} uses the \textsc{Minuit2} \cite{minuit,moneta2005developments} library for fitting data. This framework eases the analysis of muon spin rotation, relaxation, and resonance experiments by allowing the user to define all the relevant input parameters and functions for \textsc{Minuit2} in a scripting manner. At the same time, the $\mu$SR spectra are visualized utilizing the ROOT framework \cite{ROOT_Brun199781}.

\subsection{Problem description}

\begin{figure}[h]
\centering
\includegraphics[width=3in]{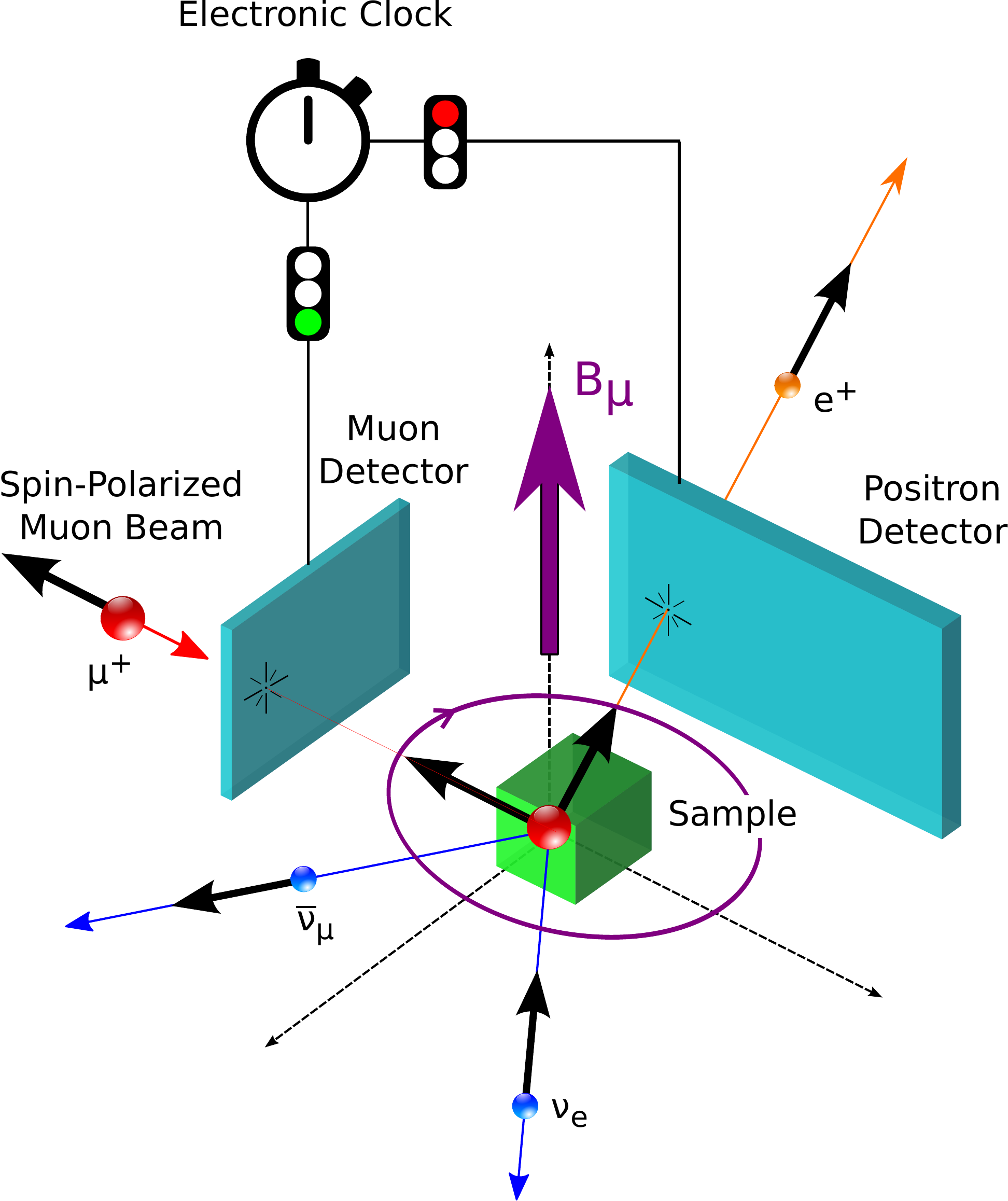}
\caption{Schematic of a time differential $\mu$SR experiment}
\label{fig:musr_experiment}
\end{figure}

The schematic of a time differential $\mu$SR experiment is shown in Figure \ref{fig:musr_experiment}. During an experiment, $\sim100$\% polarised positive muons ($\mu^+$) are implanted in a solid sample where they rapidly thermalise ($\sim 10$ ps) without noticeable polarization loss.

After the implantation the spin evolution of the muon ensemble is measured as a function of time. The evolution can be monitored by using the fact that the parity violating muon decay is highly anisotropic. During the decay an easily detectable positron is emitted preferentially along the direction of the $\mu^+$ spin. The time differential $\mu$SR spectrum takes the form:

\begin{equation}\label{eq:positron_histo}
 N^j(t, {\vec P}) = N_0^j e^{-t/\tau_\mu} [ 1 + A^j({\vec p}^j, t) ] + N_{\rm bkg}^j,
\end{equation}

\noindent where the time is measured in discrete steps $t = n \cdot \Delta t$ [$n \in \mathbb{N}_0$, $\Delta t$ the time resolution] and $j$ indexes the positron detectors. The ``physics'' of the system under consideration is described by the function $A^j({\vec p}, t)$. More details about the function $A^j({\vec p}, t)$ can be found in Ref. \cite{youanc2011muon}. The muon lifetime is given by $\tau_\mu$ and  $N_0$ gives the scale of the positron count. Lastly the constant $N_{\rm bkg}^j$ originates from uncorrelated background events. For a given positron histogram, $j$, the optimal parameter set 

\begin{equation}
 {\vec P}^j = \left\{ N_0^j, N_{\rm bkg}^j, {\vec p}^j \right\} 
\end{equation}

\noindent needs to be determined. Depending on the level of statistics of the positron histograms, the parameter set, ${\vec P}$, is determined by minimizing the $\chi^2$ function:

\begin{equation}\label{eq:chisq}
 \chi^2({\vec P}) = \sum_j \sum_n \frac{[d_n^j - N^j(t, {\vec P})]^2}{(d_{n, \rm err}^j)^2},
\end{equation}

\noindent where $d_n^j$ are the measured data points of the $j^{\rm th}$ positron detector. The theory describing the data is given by Eq.\ (\ref{eq:positron_histo}), and $d_{n, \rm err}^j$ is the estimated error of $d_n^j$ ($d_{n, \rm err}^j = \sqrt{d_n^j}$ for the Poisson distributed positron events). 

For data sets with rather limited statistics, Eq.\ (\ref{eq:chisq}) is not leading to satisfactory results. In this case the log-likelihood function


\begin{equation}\label{eq:log-likelihood}
  {\cal L}({\vec P}) = 2 \cdot \sum_j \sum_n 
               \left[ N^j(t, {\vec P}) - d_n^j \right] + d_n^j \log\left[\frac{\displaystyle d_n^j}{\displaystyle N^j(t, {\vec P})}\right]
\end{equation}

\noindent should be maximized, which leads to a much better estimate of ${\vec P}$.


With the improvements in detector technologies, it is possible to achieve higher time resolution (smaller $\Delta t$) during the experiments. This is leading to increasing sizes of data sets that need to be analyzed, and the associated minimization/maximization times are increasing drastically.

To perform the parameter fitting, \textsc{Musrfit} uses the \textsc{Minuit2} library. \textsc{Musrfit} contains the implementations of Eqs. (\ref{eq:chisq}) and (\ref{eq:log-likelihood}) while the minimization/maximization process is executed by \textsc{Minuit2}. The main, and most time consuming, part of the parameter fitting is the calculations embedded in Eqs. (\ref{eq:chisq}) and (\ref{eq:log-likelihood}) respectively. Offloading these calculations to the GPU could lead to a significant improvement in the total time needed to perform a parameter fit. In the following discussion we will use $\chi^2$ synonymous for ${\cal L}({\vec P})$ fits.

What does ``real time'' data analysis in the context of $\mu$SR mean and why is it important? $\mu$SR is a spectroscopic, accelerator based technique where measurement slots are awarded through a highly competitive proposal system. In the best case a researcher is granted a beam time slot twice a year. During these short beam periods (typically 2-4 days), all the necessary measurements need to be performed. The material classes studied by $\mu$SR are often showing a very reach and complex physics and hence it is initially hard to judge what will be the best measuring strategy in terms of available external parameters, like temperature, field, pressure etc. The online modeling of the data is crucial to conclude on an optimal measurement program. However, for some $\mu$SR instruments, currently the parameter fitting time which is needed in this modeling process is comparable to the actual measurement time. This makes it very hard to come to a clever and decisive decision how to use the beam time. Wrong decisions will force researcher to re-apply for beam time which is a waste of resources. Therefore it is crucial to reduce the fitting time to a level allowing to come to the right conclusions during online analysis. To exemplify the above stated, the numbers for the HAL-9500 instrument at the Paul Scherrer Institut can help. A typical measurement time for a given field and temperature is 2-4 hours. Robust fitting results are only available after about half of the measurement time. A single fit with the current version of \textsc{Musrfit} which  utilizes OpenMP takes about 15 min. During the first day of data taking various fitting models need to be applied and refined. This means that the full online analysis takes longer than the measurement. This is drastically improved by the DKS solution which brings the fitting times down to about 20 sec allowing to find the appropriate fitting model needed to guide the experiment successfully.

\subsection{GPU implementations}
To ease the process of adding GPU support to \textsc{Musrfit}, the Dynamic Kernel Scheduler (DKS) \cite{dks} was used. All the device specific code is developed in DKS and \textsc{Musrfit} only receives a simple interface that it can use to invoke task execution on the GPU. DKS uses CUDA or OpenCL frameworks to create the GPU code. CUDA is used to target Nvidia GPUs while OpenCL implementation is used to target devices from other vendors (Intel, AMD).

\begin{figure}[ht]
\centering
\includegraphics[height=3in, width=3in]{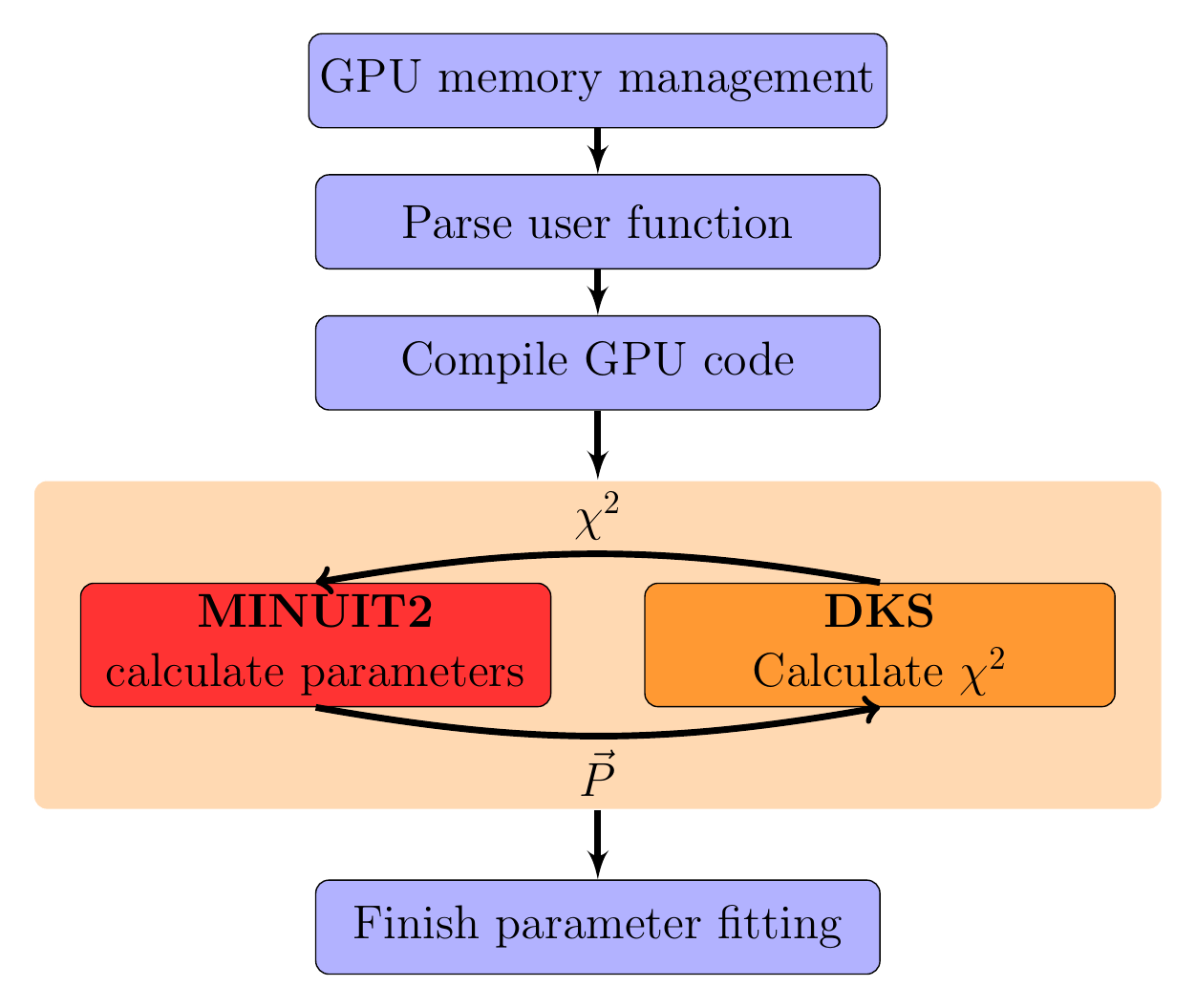}
\caption{Flow diagram of parameter fitting with \textsc{MUSRFIT} using \textsc{Minuit2} and \textsc{DKS}}
\label{fig:simplefit_sd}
\end{figure}

Using DKS, \textsc{Musrfit} allocates memory on the GPU for every data set used in the fitting and transfers the data to the device. Since the data sets do not change during the fitting, this operation can be performed only once. 

One of the most important features of \textsc{Musrfit} is the ability for users to define the theory function using the input files. A mechanism needs to be created where this user defined function can be passed to the GPU at run-time and used in the kernel code. To handle this problem, run-time compilation was used. The user defined function is parsed by \textsc{Musrfit} from the input files and passed to DKS where a CUDA or OpenCL device function is created to be used in GPU kernels. This process is described in more detail in section \ref{sec:userfunc}. When the new GPU program is created and compiled by DKS, \textsc{Musrfit} begins the process of minimizing the $\chi^2$ value by invoking the CUDA or OpenCL kernels to calculate the $\chi^2$ value and using \textsc{Minuit2} to fit the parameter set. The sequence diagram of this process is shown in Figure \ref{fig:simplefit_sd}.

\subsubsection{User-defined functions}
\label{sec:userfunc}
To allow users to define functions, the GPU code must be created at run-time. For OpenCL this is the standard execution method, while for the CUDA framework the CUDA run-time compilation library \cite{cuda-rt} was used. \textsc{Musrfit} parses the user input file to get the user defined function and creates a string with a C++ mathematical expression. This expression can use standard C++ mathematical operators and functions, and in addition it is able to utilize a set of predefined functions which are commonly used in $\mu$SR field. CUDA implementations of exponential and Gaussian distribution functions are shown below in code sample \ref{code:predefinedfunc}. A full list of available predefined functions is listed in \textsc{Musrfit} user guide \cite{suter2012musrfit}.

 \begin{figure}[ht]
 \centering
  \begin{lstlisting}[frame=single, caption={CUDA examples of predefined functions that can be used to create the user function.}, label={code:predefinedfunc}]
__device__ 
double @@se@@(double @t@, double @lambda@) {
  return exp( -lambda*t );
}

__device__ 
double @@ge@@(double @t@, double @lambda@, double @beta@) {
  return exp( -pow(lambda*t, beta) );
}

__device__ 
double @@sg@@(double @t@, double @sigma@) {
  return exp( -0.5 * pow(sigma*t, 2) );
}

__device__ 
double @@stg@@(double @t@, double @sigma@) {
  double @sigmatsq@ = pow(sigma*t,2);
  return (1/3) + (2/3)*(1 - sigmatsq) * exp(-0.5 * sigmatsq);
}
  \end{lstlisting}
 \end{figure}                    

The mathematical expression can use the parameter array to access parameter values and the function array to access precomputed function values. The function array is a convenience feature for the user. A subset of the parameter array is data set specific. In order to keep the mathematical expression compact, an indirect addressing of these parameters is needed. This is accomplished with the map array. For more details see Ref.\cite{suter2012musrfit,musrfit2015userManual}. An example of a created user function for use in CUDA kernels is shown in code sample \ref{code:userfunc}.

\begin{figure}[ht]
  \centering
  \begin{lstlisting}[frame=single, caption={Example of parsed user defined function ready for compilation.}, label={code:userfunc}]
__device__ 
double @@fTheory@@(double @t@, double *@p@, double *@f@, int *@m@) 
{
  return p[m[0]] * sg(t,p[m[1]]) * tf(t,p[m[2]],f[m[3]]);
}
  \end{lstlisting}
\end{figure}

After \textsc{Musrfit} has created the string containing the mathematical expression of the user defined function, it is added to the string containing the CUDA program. The CUDA program consists of a user defined function definition, predefined functions, and the kernel for $\chi^2$ calculation. This newly created program is compiled at run-time and used by DKS to evaluate the $\chi^2$ of a given data set.

\subsubsection{Computing $\chi^2$}
The most time consuming part of the parameter fitting is the calculation of the $\chi^2$ function for each data set. This calculation can be easily parallelized and therefore is an ideal candidate to offload to the GPU. The CUDA kernel to compute the $\chi^2$ value creates a thread for each data point in a data set. Shared memory is used to store parameter, function, and map values since these values are accessed multiple times by each thread. Using the new parameters, functions, and maps for the data set, the theory function is evaluated at each point and the $\chi^2$ value at that point is calculated and stored in a temporary allocated global memory array. After the kernel completes the calculation of $\chi^2$ for each individual data point, all these values are summed up using CUBLAS \cite{cublas} to get the $\chi^2$ value of the whole data set. This process is repeated for every data set used in the calculation.

\subsection{Results}

The tests of the parameter fitting were run on two systems. First system was equipped with two Intel(R) Xeon(R) CPU E5-2609 v2 processors and one Nvidia Tesla K40c GPU. The second system was equipped with two Intel(R) Xeon(R) CPU E5-2690 v3 processors. \textsc{Musrfit} parallelizes CPU code using OpenMP so the performance of the fitting using this implementation, with 8 threads, was chosen as the baseline. To test the CUDA and OpenCL performance, the same example was run on the GPU using both of these frameworks. Another benchmark was run with OpenCL using the CPU as the target device on the first machine.

For the tests, a typical muon polarisation function was chosen to determine the magnetic shift of a para-/diamagnetic material \cite{youanc2011muon}. It is given by:

\begin{equation}\label{eq:asymmetry}
 A^{j}({\vec p},t) = A_0^j \exp\left[-\frac{1}{2} (\sigma t)^2\right] \, \cos(\gamma_\mu B t + \phi^j),
\end{equation}

\noindent where $j = 1$ to 16, where 16 is the number of positron detectors in this example. $A_0^j$ is the asymmetry of each positron detector, $\sigma$ is the depolarisation rate of the muon spin ensemble, $\gamma_\mu$ is the gyromagnetic ratio of the muon, $B$ is the magnetic induction at the muon stopping site, $t$ is the time, and $\phi^j$ is the phase of the initial muon spin in respect to the positron detector.

      

 \begin{table}[ht]
   \caption{Parameter fitting with $\chi^2$ function running on the GPU. The given time is for the execution of the \texttt{minimize} command of Minuit2 \cite{moneta2005developments}.}\label{tab:minuit2-results}
   \centering
   \normalsize{
     \begin{tabular}{c | c | c | c | c}
       Data size & Iter. & Device & Time (s) & Speedup\\
       \hline\hline

       \multirow{3}{*}{16$\times$85320} & \multirow{3}{*}{8833} & E5-2609 & 290 & \\
       & & E5-2690 & 226 &\\
       & & Tesla K40c & \textbf{11} & \textbf{\textcolor{red}{$\times$20 to $\times$26}}\\
       \hline
       \multirow{3}{*}{16$\times$106650} & \multirow{3}{*}{8538} & E5-2609 & 351  &\\
       & & E5-2690 & 274 &\\
       & & Tesla K40c & \textbf{11.5} & \textbf{\textcolor{red}{$\times$23 to $\times$30}}\\
       \hline
       \multirow{3}{*}{16$\times$142200} & \multirow{3}{*}{9319} & E5-2609 & 508 &\\
       & & E5-2690 & 396 &\\
       & & Tesla K40c & \textbf{13.8} & \textbf{\textcolor{red}{$\times$28 to $\times$36}}\\
       \hline
       \multirow{3}{*}{16$\times$213300} & \multirow{3}{*}{8052} & E5-2609 & 654 &\\
       & & E5-2690 & 513 &\\
       & & Tesla K40c & \textbf{15.1} & \textbf{\textcolor{red}{$\times$33 to $\times$43}}\\
       \hline
       \multirow{3}{*}{16$\times$426601} & \multirow{3}{*}{6313} & E5-2609 & 1015 &\\
       & & E5-2690 & 798 & \\
       & & Tesla K40c & \textbf{17.9} & \textbf{\textcolor{red}{$\times$44 to $\times$56}}
      
     \end{tabular}
   }
   \label{table:chisquare}
 \end{table}

The results of these tests are shown in the Table \ref{table:chisquare}.\ The results show that for the chosen test function, the total execution time of the parameter fitting can be improved by around $\times$40 to $\times$50 on the GPU depending on the size of the problem. 

\begin{figure}[ht]
\centering
\includegraphics[height=2.5in, width=3in]{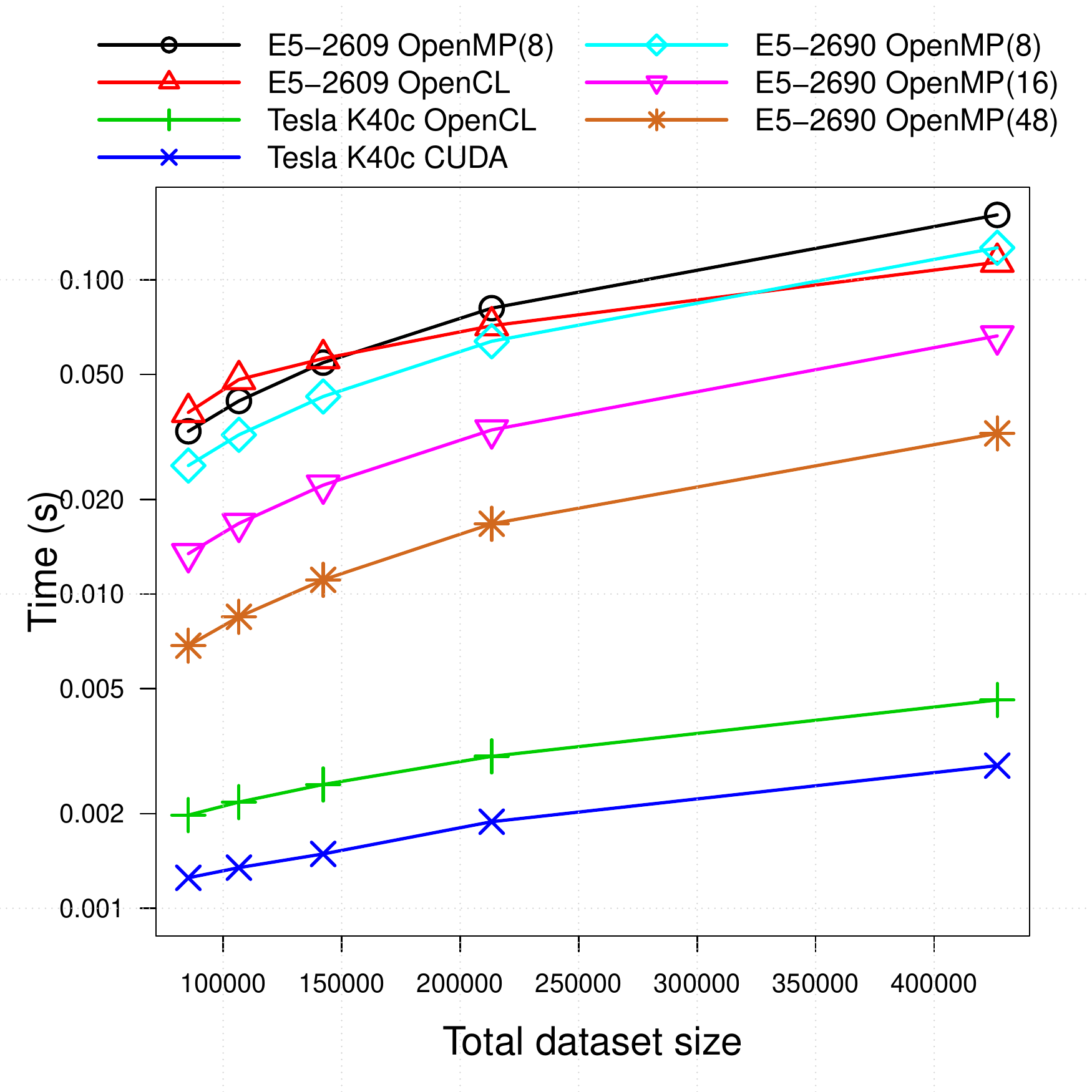}
\caption{Parameter fitting with $\chi^2$ function running on the GPU. The time is shown for the execution of one iteration of the \texttt{minimize} command of Minuit2 \cite{moneta2005developments}.}
\label{fig:chisquare-results}
\end{figure}

The OpenCL implementation of parameter fitting in DKS allows the use of other accelerator devices to speed up the calculations. This makes the application more portable and more accessible to users.\ The results of OpenCL tests are shown in Figure \ref{fig:chisquare-results}. This figure also shows OpenMP results when using up to 48 CPU cores to run the fitting on the second test system equipped with two Intel(R) Xeon(R) CPU E5-2690 v3 processors each consisting of 24 virtual cores with hyperthreading enabled.

\add{The performance increase of \textsc{Musrfit} with GPUs using CUDA or OpenCL allows for data analysis of $\mu$SR experiments to be done real-time, which will lead to increased efficiency of the experiments. Furthermore the use of DKS layer in \textsc{Musrfit} will allow to ease the maintenance of the GPU code to adjust for new GPU architectures or introduction of new devices such as Intel MICs or FPGAs.}

\goodbreak
\section{Image reconstruction and\\ analysis}
\label{sec:recon}
The SAFIR (Small Animal Fast Insert for MRI) project is developing a fast PET insert 
for a pre-clinical MRI (Magnetic Resonance Imaging) system for dynamic in vivo PET-MRI studies with 
excellent temporal resolution. This requires tomographic image reconstruction 
followed by image data analysis adapted to the conducted study. 
While under idealized assumptions, the image can be obtained 
analytically by a filtered inverse Fourier transform. Modeling the 
system details and irregularities results in better images. 
However, this second approach involves the manipulation of huge matrices.
Therefore typically iterative 
image reconstruction algorithms are applied, which still constitute
a significant computational burden. 

Image analysis, such as feature finding, is computationally time intensive as well. 
Moreover, in dynamic studies sequences of dozens of images need to be reconstructed 
and analyzed. 
In particular the aim is to reconstruct one image about every 5 s. Thus 60 images need to be reconstructed for a typical acquisition of 5 minutes, each comprising 5 s worth of data.
Computation times of hours to days would be required for 
data acquired in a few minutes. Speeding up the computation is 
therefore of prime importance. 
A first big improvement would be the reconstruction of the image series on a time-scale of one hour.
Ultimately, a quasi-online visualization of the process dynamics would be very beneficial 
to control and optimize the experiments requiring to reconstruct one image within 5s. 

\add{There are several articles in the literature that describe the efforts of accelerating PET image reconstruction codes using GPUs \mbox{\cite{Cui2011,Herraiz2011,Pratx2011}}. The algorithm described in this work uses list-mode data for image reconstruction and follows a similar approach as proposed in \mbox{\cite{Cui2011}} and \mbox{\cite{Pratx2011}}. The results obtained in previous works show that PET image reconstruction is a good algorithm for GPU acceleration and would greatly benefit the reconstruction algorithms used in SAFIR project.}

\subsection{Image reconstruction}

\begin{figure}[ht]
  \centering
  \includegraphics[height=2.5in, width=2.5in]{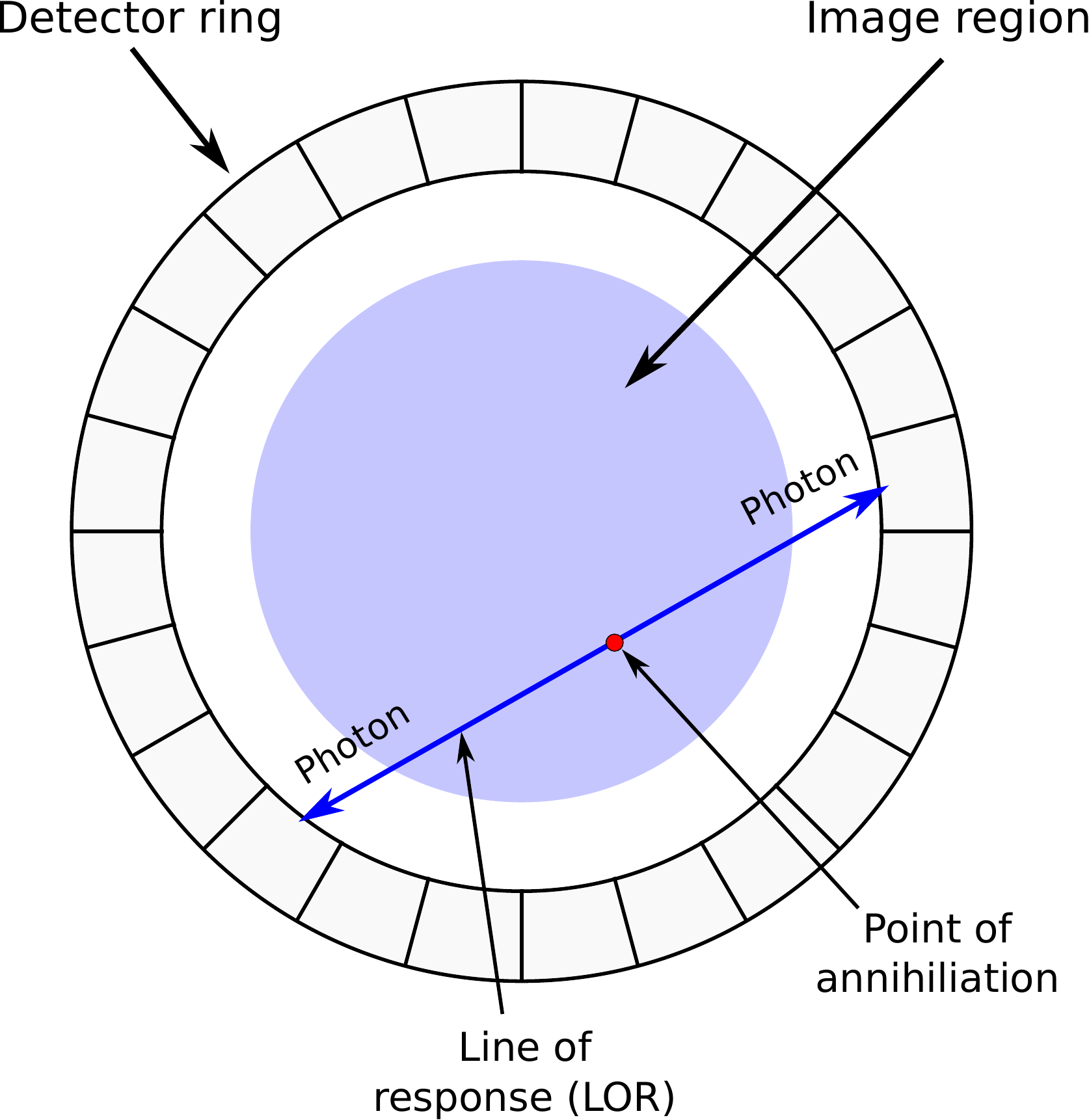}
  \caption{PET imaging basic principles.}
  \label{fig:pet_basic}
\end{figure}

In PET image reconstruction, the goal is to find the source activity distribution 
in the object to be studied. The activity distribution is found from the projection measurements 
of the set of coincident detector pairs, which form the whole PET scanner. 
In PET imaging, a positron emitting radiotracer is used.
The positron annihilates producing two back-to-back photons with 
511 keV energy. These are measured,
typically with a set of cylindrically arranged detectors surrounding the source.
A schematic sketch of one ring of such an arrangement is shown in figure \ref{fig:pet_basic}.

When within a short time window two detectors each register a photon, 
it indicates that an annihilation event has occurred on the line joining
the two detectors (line-of-response, LOR).
This is a measurement of projections, because the number of counts that have happened
on a given line can be approximated as the line integral of the activity distribution
along this line.

The list of all these coincidence events (listmode data) can directly be
used to reconstruct the image.
A general description of PET image reconstruction can be found in \cite{recon_overview}.

Consider a discrete tracer distribution $\vec{f} = (f_1,\dots,f_J)^T$ on a 3D Cartesian grid, 
where the index labels the volume elements called voxels and $J$ is the number of voxels.
The coincidence measurement yields an estimation of the mean counts 
$\vec{\bar{y}} = (\bar{y}_1,\dots,\bar{y}_I)^T$
per detector pair $i$, where $I$ is the number of possible
pairs.

Using a linear model for $\bar{y}_i$, the relationship between
tracer distribution and mean measured counts can be written as

\begin{equation}
\bar{y}_i = \sum_{j=1}^J a_{ij} f_j + n_i,
\end{equation}

\noindent where the matrix $\mathbf{a}$ is called the system matrix and $\vec{n}$ is a noise term.
PET reconstruction aims at solving the inverse problem, i.e. finding $\vec{f}$
given $\vec{\bar{y}}$. Inverting the matrix $\mathbf{a}$ is computationally unaffordable
due to its size. Therefore, iterative approaches are employed.

The stochastic nature of the radioactive decay, together with the detection
of the events, can be modeled with a Poisson process. The likelihood function is

\begin{equation}
p(\vec{\bar{y}} | \vec{f}) = \prod_{i = 1}^I p(y_i|\bar{y}_i) = \prod_{i = 1}^I e^{-\bar{y}_i} \frac{\bar{y}_i^{y_i}}{y_i!},
\end{equation}

\noindent where $y_i$ is the actually measured number of counts in the $i$-th line-of-response (LOR).

The Bayes factors for the inversion of the conditional probability are neglected.
Finding the minimum of the likelihood function with respect to $\vec{f}$ yields
an iterative formula for the distribution

\begin{align}
f_j^{k+1} &= \frac{f_j^k}{\sum_{i=1}^I a_{ij}} \sum_{i=1}^I a_{ij} \frac{y_i}{\bar{y}_i^k}, \\
\bar{y}_i^k &:=  \sum_{j=1}^J a_{ij} f^k_j + n_i.
\label{eq:fwdproj}
\end{align}

This algorithm can be rewritten for listmode processing

\begin{align}
f_j^{k+1} =& \frac{f_j^k}{\sum_{i=1}^I a_{ij}} \sum_{l=1}^L \sum_{i=1}^I \delta_{i,c(l)} \frac{a_{ij}}{\bar{y}_i^k}
\label{eq:lm_recon_iter}
= \frac{f_j^k}{\sum_{i=1}^I a_{ij}} \sum_{l=1}^L \frac{a_{c(l),j}}{\bar{y}_{c(l)}^k}, \\
\notag c(l):=&\text{index of detector pair} \\
&\text{corresponding to }l\text{-th listmode event}, 
\end{align}

\noindent where $L$ is the number of listmode events and $\delta_{ij}$ is the Kronecker delta. 
\add{The matrix element $a_{c(l),j}$ describes the probability of detecting an annihilation from 
the $j$-th voxel in the $c(l)$-th LOR, in which the $l$-th coincidence event was detected.}

Equation \eqref{eq:fwdproj} implicitly reappears in the denominator 
of the last term of \eqref{eq:lm_recon_iter} and is called forward 
projection because it constitutes a map from the spatial activity distribution
to the number of counts in the LORs. The sum over $l$ in  \eqref{eq:lm_recon_iter}
is called backward projection because it constitutes a map from
the set of LOR count values to the spatial activity distribution.

The matrix elements $a_{ij}$ can be estimated using an adapted raytracing algorithm.
To reduce the computation effort, the LOR's predominant direction of propagation
is determined to be along the x- or y-axis and the planes perpendicular to that axis,
through the voxel centers, are considered. For each plane, the intersection point 
$\vec{p}=(p_x,p_y,p_z)^T$ of the LOR $i$ with
that plane is determined and the voxel $j$ with center coordinate
$(v_{jx},v_{jy},v_{jz})^T$, containing this intersection point is identified. 

Without loss of generality, let the predominant direction be along the x-axis.
In each plane, the matrix element is calculated for the voxel $j$ and its
three neighbors $j',j'',j'''$ in the positive y- and z-directions. 
The matrix element $a_{ij}$  is approximated to be related to
the distance of the voxel to the intersection point in the following way

\begin{equation}
a_{ij} \approx m_d - \sqrt{(p_y-v_{jy})^2 + (p_z-v_{jz})^2 },
\end{equation}

\noindent where $m_d$ is the matrix distance factor. The matrix distance factor 
acts as a weight for the 
influence of the distance of the voxel to the intersection point to
the system matrix element.

\subsection{Image analysis}
An important task for the envisaged research is to identify a small spot,
with a volume of about 5 to 10 $\rm mm^3$, in a rodent brain with enhanced activity. This spot needs
to be identified in non-uniform background and with the enhanced activity of the order
of 20\% compared to its normal state.
The stochastic nature of the data and the 
relatively low number of counts in the few second time intervals 
results in large variance of the reconstructed activity concentration. 
It is therefore important to distinguish true features from 
fluctuations with quantifiable significance.

As explained above, the goal is to find the relative activity increase
in a region over some normal (background) activity. Hence 
the excess $E$ and its standard deviation $\Delta E$ are defined

\begin{align}
E &= \frac{S-B}{B}, \\
\Delta E &= \frac{S}{B} \sqrt{\left( \frac{1}{S} + \frac{1}{B} \right)}, 
\end{align}

\noindent where $B$ is the background activity and $S$ is the activity in the
region of interest including the background. For simplicity, two 
concentric spheres are used, the smaller (inner) one representing the signal region
and the larger, with the volume of the smaller sphere cut out, representing 
the background region. 
The significance of the excess can be expressed in terms of its standard 
deviations and a threshold can be used to separate true from random signals.

The image is processed by displacing the center of the sphere
into the center of each voxel. Applying a threshold to the transformed
image allows to locate features of a certain significance. 

\subsection{GPU Implementation}

For image reconstruction, the most time consuming parts of the algorithm are the forward and backward projections. Both of the projections loop through the projection lines and either accumulate image data along the line (forward projection) or distribute projection values into the image data along the same lane (backward projection). Every line in the list can be processed independently so this problem can be parallelized to take advantage of the computational resources available on GPU. However, there are several challenges that must be considered for the GPU algorithm to achieve the desired performance:
\begin{itemize}
\item{Some lines in the list do not require processing and different predominant line directions require alternative processing which results in a large thread divergence;}
\item{Each line requires a different set of voxels from the image resulting in random memory access;}
\item{Multiple lines need to update the same voxel in the image and thus requires atomic operations.}
\end{itemize}

For image analysis, the most time consuming part is the calculation of the average value and the standard deviation of the voxels inside a sphere. 
Two spheres are placed at the source location. First the average value and standard deviation of voxels that are inside the smaller of the two spheres (source value) are calculated. The second part of calculations finds the same values for voxels that are outside of the smaller sphere, but inside the larger sphere (background value).

Since the spheres are placed at the center of each voxel, this can be parallelized on the GPU by every thread calculating the average value for a different sphere.

To add GPU support to host application CUDA kernels for forward projection, backward projection, source calculation and background calculations were implemented in DKS. The host application was updated to use DKS instead of CPU implementation when a GPU is available. The host application uses DKS interface to invoke memory management, data transfer and kernel calls while all the temporary memory, kernel details and kernel launch parameters are handled by DKS. A DKS call sequence is presented in Code listing 1.

\subsubsection{Forward and backward projections}
The forward projection in DKS is implemented as two kernel calls. The first kernel call calculates the predominant direction of each line, and assigns a label to it:

\begin{itemize}
\item 0 - line does not need to be processed
\item 1 - predominant direction in the x plane
\item 2 - predominant direction in the y plane
\end{itemize}

\noindent Once the predominant direction of the line is known, the Thrust sort by key function is used, sorting the lines according to its direction. In this way we minimize the thread divergence in the kernel call that performs the forward projection.

\begin{figure}[ht]
\centering
\includegraphics[height=2.5in, width=2.5in]{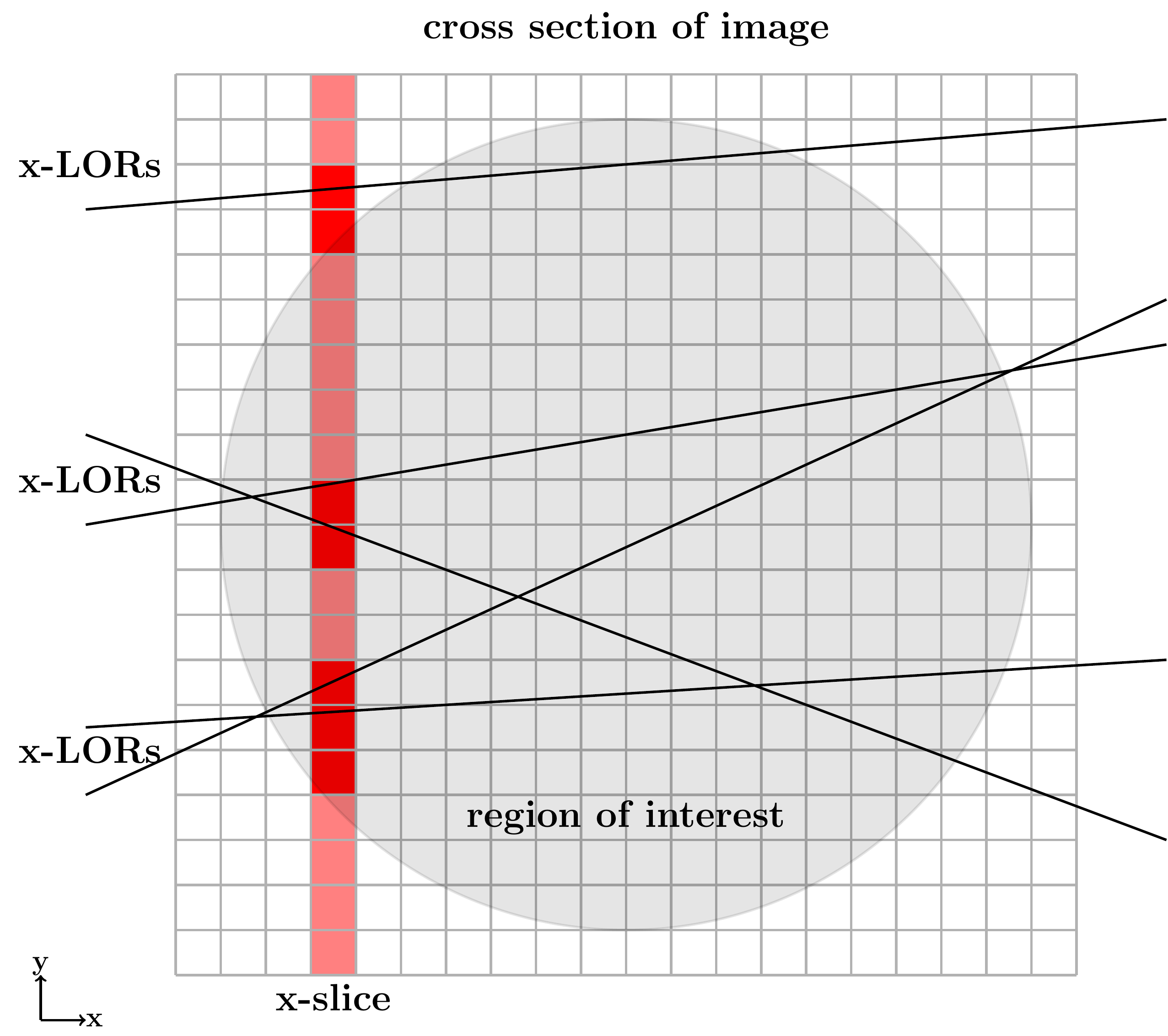}
\caption{Cross section of the image showing LORs, with predominant direction in the x plane, and a slice of the image along this direction being processed.}
\label{fig:slice}
\end{figure}

After the lines are sorted, the forward projection processes the image in slices along the predominant direction as shown in Figure \ref{fig:slice}. Whether the line requires any values from a slice is determined by the position and angle of the line. This process is the same for both forward and backwards projections. For each slice the position where the line crosses the slice is determined, if this position is inside the image region the values are loaded from global memory (forward projection) or global memory is updated with the correction value (backward projection).

The assigned label allows us to process all the lines with a single kernel call and avoid repeated calculations on how the line should be processed. After the lines are sorted, they are grouped by the predominant direction. The divergence of threads within the warp will be very minimal.

Since the lines require very few values from a slice and not all the lines use every slice, each thread checks if any voxels from the slice are needed and only then perform a load from global memory. This will result in an un-coalesced global memory access since lines in the same warp are accessing random voxels in the image. Since only a few of the slices are used for each line, and only a few values from each slice are used, this results in a better performance than loading all of the voxels in the slice in shared memory.

Backward projection takes the correction value calculated for each line and distributes it back to the voxels along this line. During backward projection, the same sorted list of lines is used to tackle thread divergence. The main bottleneck for the backward projection is the need for multiple threads to update the same voxels because different lines can cross the same positions in the image. To avoid race conditions when multiple threads need to write to the same global memory address CUDAs atomic operations are used.  

DKS calls are inserted in the host application to offload tasks if a GPU device is present. Using the DKS, the  host application allocates memory on the device and transfers data from the host, holding voxel positions, voxel values, detector pair list and detector positions. In addition memory is allocated on the GPU to hold the correction values for each detector pair calculated by forward projection and corrected voxel values calculated by background projection.
After memory allocation and data transfer the host application loops trough the set reconstruction iterations and uses DKS to call forward and backward projection kernels on the GPU. Every iteration requires a read of corrected voxel values from the GPU, and since the final processing of the image is done by the host application, before the next iteration new voxel values are written back to the GPU. Each iteration also requires a write of list of detector pairs used for reconstruction. Since after every iteration half of the detector pairs are discarded this list needs to be updated and resorted before every forward projection. The example code of host application and DKS integration for image reconstruction is shown in the code sample \ref{code:dksrecon}.

\begin{figure}[ht]
\centering
\begin{lstlisting}[frame=single, caption={Example code of DKS interface integrated in the host application for image reconstruction.
Initialization of DKS, memory allocation and deallocation is similar to the Code sample 1.}, label={code:dksrecon}]
for (int iter = 0; iter<num_of_iteration; iter++) 
{
  //transfer image data to GPU every time step
  dksbase.writeData<float>(*image_gpu, *recon_image_host, image_size);

  //calc forward projections on the GPU
  dksbase.callForwardProjection(*line_correction, *image_gpu, *list_detectors, *detector_position, *image_position, event_number);
  
  //calc backward projections on the GPU
  dksbase.callBackwardProjection(*line_correction, *image_correction, *list_detectors, *detector_position, *image_position, event_number, image_size);

  //read recon_image_3d_corrector form GPU
  dksbase.readData<float>(*image_correction, *recon_image_host, image_size);

  //final processing of the reconstructed image
  //output operations

  //remove half of detector pairs
  event_number /= 2;
  dksbase.writeData<ListEventData>(*list_detectors, *list_detectors_host, event_number);
}
\end{lstlisting}
\end{figure}

\subsubsection{Source and background calculation}
\begin{figure}[ht]
\centering
\includegraphics[height=2.5in, width=2.5in]{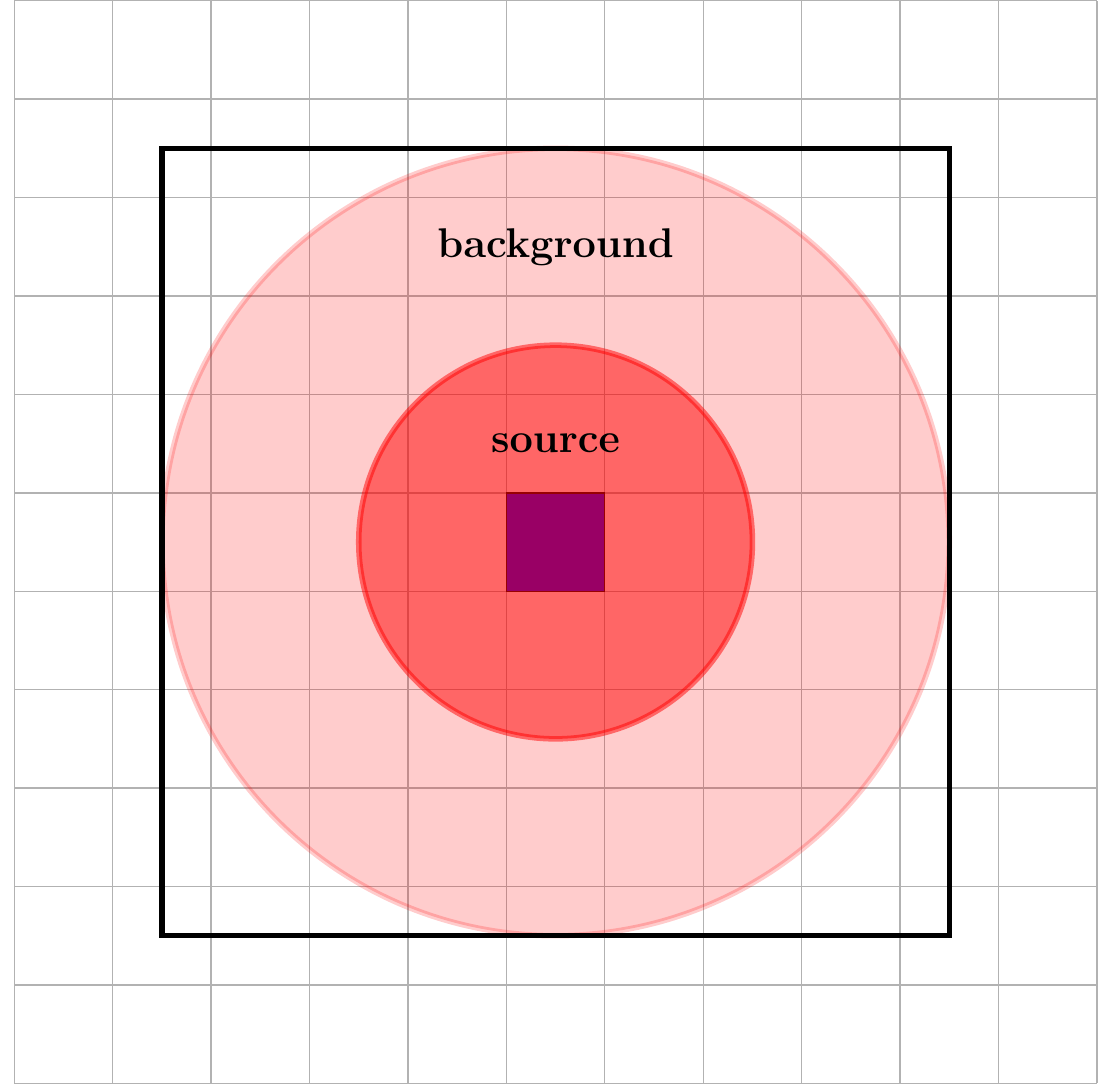}
\caption{2D representation of sphere placement at the voxel position for source and background calculation.}
\label{fig:sphere}
\end{figure}

Source and background calculations are separated in two kernels. To calculate the source values at each voxel position, every thread places a sphere with the center at this voxel. Then knowing the diameter of the sphere, the position of the voxel, and the size of each voxel, a box is calculated that contains this sphere. This process is illustrated in Figure \ref{fig:sphere}.

After the box is formed the thread loops through the voxels in this box and calculates the average value and standard deviation using only voxels that lie inside the sphere. The box is necessary to minimize the number of voxels that each thread has to process. This approach requires a lot of global memory access. Shared memory usage could be explored to improve the performance of the kernel, since voxels used by threads in the same thread block are overlapping.

To calculate source and background values on the GPU, the host application first needs to allocate memory on the device and transfer data for voxel positions and voxel values in the image. Then memory is allocated to hold temporary values for average values and standard deviation at each sphere. When all the necessary data is transferred to the device, kernels to calculate source and background values are invoked trough DKS. In the final stage of the algorithm the host application reads the average and standard deviation values for each sphere from the GPU and performs final signal to noise ratio calculations. As in the case of forward and backward transformation the host application is responsible when memory allocation and data transfer is scheduled, in order to ensure that the host application can access the data from GPU when needed, but DKS handles all the device code details and kernel launch parameters.








\subsection{Results}
The tests for image reconstruction and analysis where performed on the same two systems as the parameter fitting tests. The first system uses Intel(R) Xeon(R) CPU E5-2609 v2 processor and Nvidia Tesla K40c GPU, while the second system uses Intel(R) Xeon(R) CPU E5-2690 v3 CPU. The CPU implementation of reconstruction and analysis is not parallelized, so the single CPU core performance was chosen as the baseline performance.

To test the GPU performance, simulated data were used. The data were generated 
using GEANT4 \cite{Agostinelli,Allison} for an idealized scanner made from 91 rings of 180 detectors. 
The detector crystals are 2.0 mm x 2.0 mm and are 12.0 mm long in the radial direction. 
The pitch between adjacent detectors in a ring, as well as between the rings, 
is 2.2 mm. Simulations of a Derenzo \cite{Budinger} type phantom were performed:
six groups of spheres 
with different diameters (1.0 mm, 1.2 mm, 1.6 mm, 2.4 mm, 3.2 mm, and 4.0 mm) 
were embedded into a rat phantom. The rat phantom was implemented as a high density polyethylene cylinder,
with a length of 150 mm and a diameter of 50 mm. The simulation was performed 
for one second with 500 MBq distributed evenly over the spheres volume.
This corresponds to 1.42 MBq/$\rm mm^3$ and zero activity in the rat phantom. 
Reconstruction and image analysis were performed using the algorithms as described above.

The reconstructed image size was 90x90x50 voxels with a voxel size of 0.7 mm x 0.7 mm x 0.7 mm, and the reconstruction 
was performed using $13,901,607$ \add{coincidence events}. The reconstruction starts by using 
all of the available \add{coincidence events} and performs forward and backward projections 
for 15 iterations. 


\begin{table}[ht]
  \caption{Performance of image reconstruction and analysis example.}
  \centering
  \normalsize{
    \begin{tabular}{c | c | c | c | c}

      Device & Recon & Speedup & Analysis & Speedup\\
      \hline
      \hline
      E5-2609 v2 & 800s & & 8.8s & \\
      E5-2690 v3 & 599s & & 5.9s & \\
      Nvidia Tesla K40c & 14 s & \textbf{\textcolor{red}{$\times$57} } 
      & 2.7s & \textbf{\textcolor{red}{$\times$3}}\\ 
    \end{tabular}
  }
  \label{table:recon}
\end{table}

When performing image analysis, the example performs two separate types of analysis. The first analysis places the spheres at previously defined source positions and the second analysis places the spheres at every voxel that lies inside the image region.

\begin{figure}[ht]
  \centering
  \includegraphics[height=3in, width=3in]{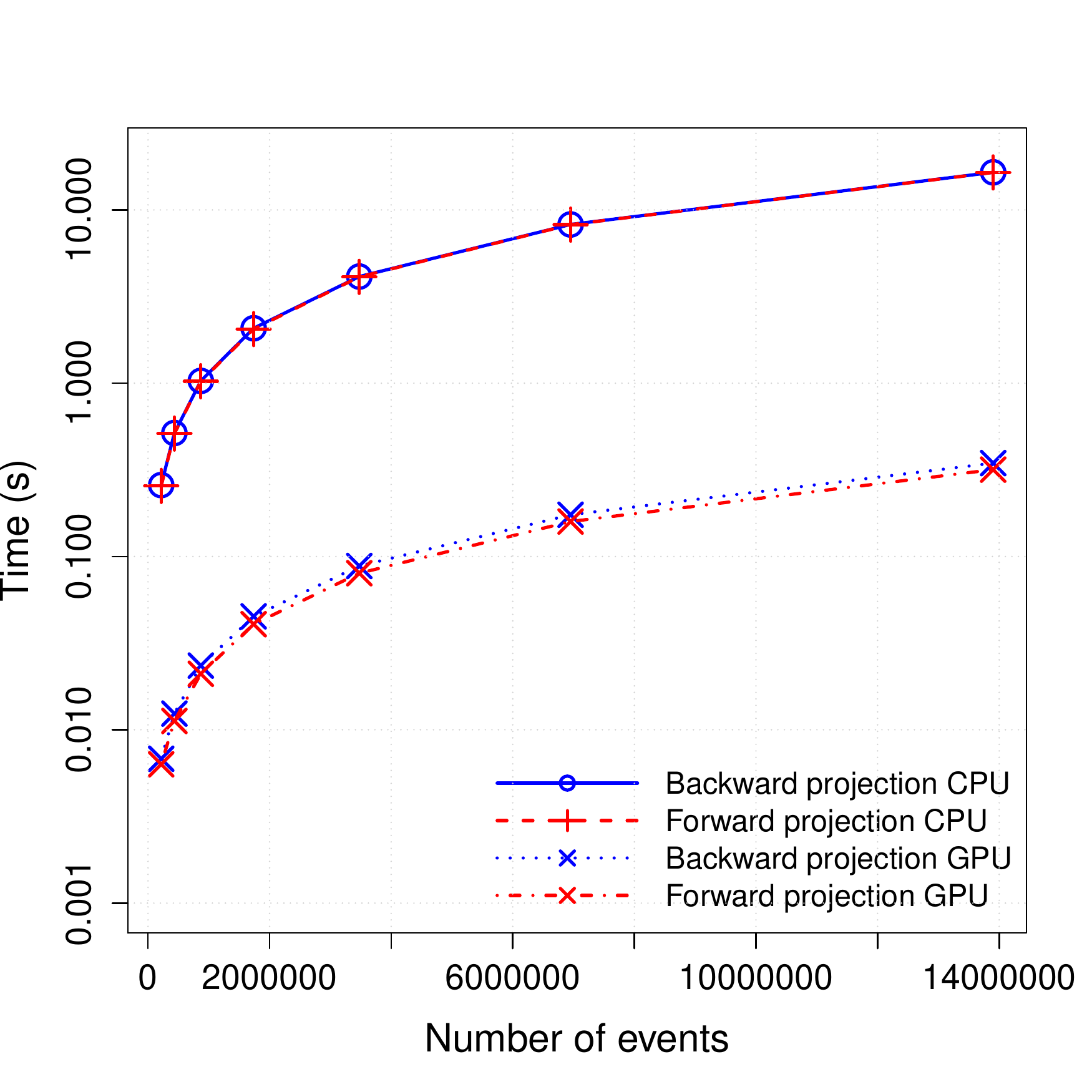}
  \caption{Execution time for forward and backward projections. Run on Intel(R) Xeon(R) CPU E5-2690 v3 and Nvidia Tesla K40c}
  \label{fig:projections}
\end{figure}

The results of the reconstruction for 1s of data and the analysis are shown in Table \ref{table:recon}. As can be seen from the results, the implemented GPU version cuts the execution time from almost 15 minutes to around 15 seconds. The time represented in the table shows the total execution time of the reconstruction algorithm, including the input and the output operations. For image analysis input and output operations are excluded from the benchmarking, because they take more that 50\% of total analysis time. For the image analysis, the diameter of inner and outer spheres are chosen to be 2mm and 4mm. The performance of individual kernels, for image reconstruction, offloaded to GPU are shown in Figure \ref{fig:projections}. The results in Figure \ref{fig:projections} illustrate the execution time for forward and backward projections using different numbers of lines for image reconstruction on a CPU and a GPU.

\begin{figure}[ht]
  \centering
  \includegraphics[height=3in, width=3in]{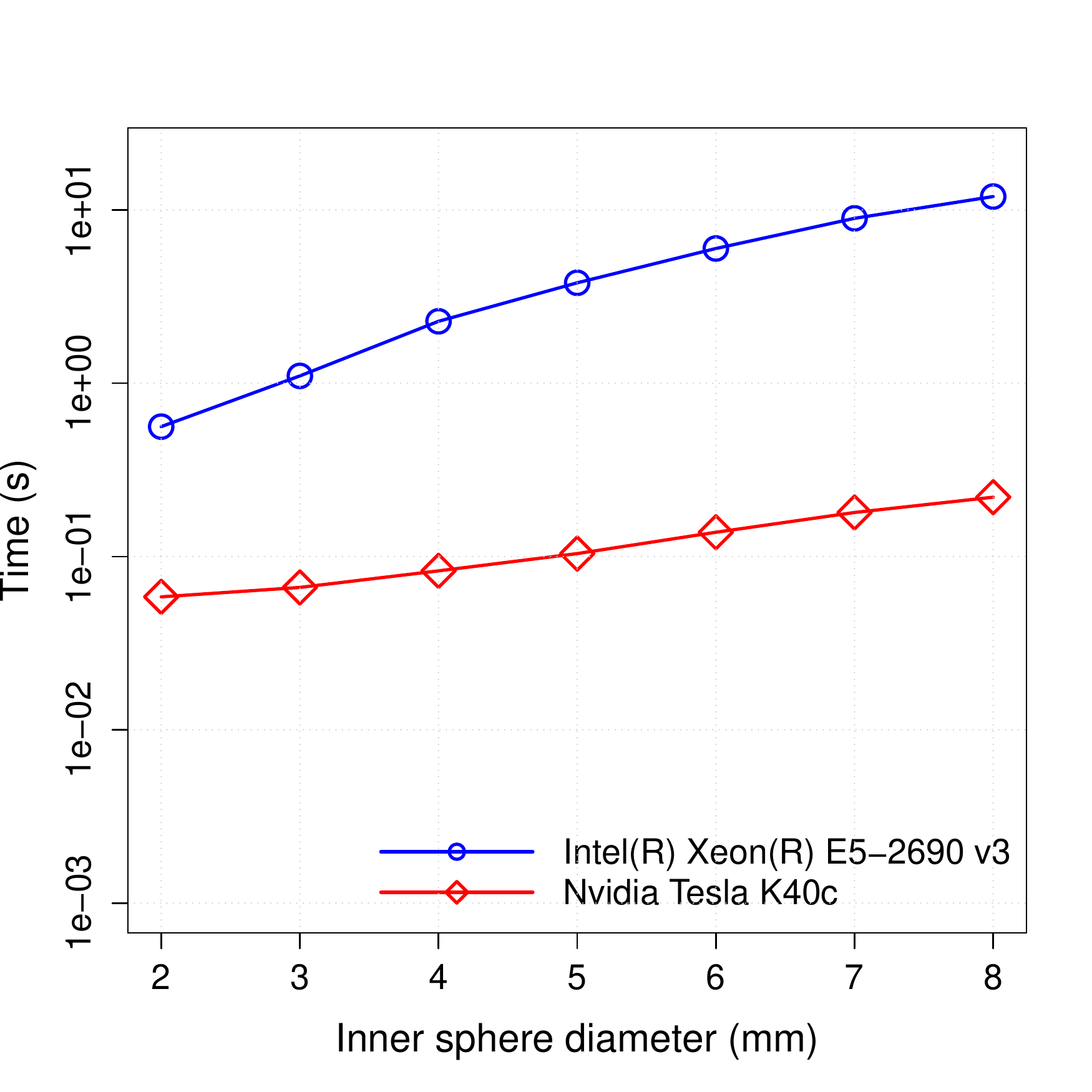}
  \caption{Calculation of source and background values with different sphere diameters.}
  \label{fig:analysis_diameter}
\end{figure}

To test the GPU performance and scaling of kernels used for image analysis, tests were repeated with different sphere sizes. Figure \ref{fig:analysis_diameter} shows the execution time of calculating source and background values at each voxel position with different sphere diameters. The diameter of the outer sphere is always twice the diameter of the inner sphere.

The computation time for the image reconstruction depends on the
total number of detected events and on the number of iterations, 
but is independent of the true image and the physical scanner used 
to obtain the data. 
The computation time of the image analysis will scale 
with the number of voxels and the size of spheres, as 
can be seen in Figure \ref{fig:analysis_diameter}. However the computed time  is independent 
of the content of the 
image and of how the image was actually obtained. Therefore, the results 
of the examples are representative for all other possible PET systems,
applying similar reconstruction and feature finding. 
In the given application using list-mode processing, the reconstruction time scales linear with the amount of data. Thus reconstructing an image of 5 s using the GPU requires a total of \add{72.7} s, and the reconstruction of a series of 5 minutes of data into 60 images would require \add{1 h and 12.7 min}. This is a significant improvement compared to the \add{\midtilde66} hours required in the CPU case and already close to the first objective to reconstruct within the order of one hour.

\section{Conclusions}
\label{sec:conclusions}
The Dynamic Kernel Scheduler (DKS) was used to enable GPU usage in two different scientific applications. Both use cases
are aimed at two separate problems, data analysis of $\mu$SR experiments and PET image reconstruction/analysis respectively. 

Significant speedups were reached, that allows, for the first time, near real-time data analysis in both use cases. 

Due to the higher level of abstraction achieved with DKS, the application development and maintenance will be simpler and independent of the fast moving hard- and software technology. 

This is software investment protection, which is important, and will become an even bigger issue in the present and future trend towards more heterogeneous hardware and software environments.

Future
work will include the use of multiple GPU's to cover larger data sets, and explore an efficient OpenCL implementation with the
goal to target a wider range of hardware. Auto-tuning on the level of optimal kernel launch parameters is planed for future releases of DKS.



%

\section{References}
\bibliographystyle{elsarticle-num}
\bibliography{dksbib}{}  
%
%

\end{document}